# 业余天文学在中国：现状与未来

叶泉志[1, 2]

[1] Division of Physics, Mathematics and Astronomy, California Institute of Technology, Pasadena, CA 91125, U.S.A.
[2] Infrared Processing and Analysis Center, California Institute of Technology, Pasadena, CA 91125, U.S.A.

**摘 要**：公众科学(citizen science)指的是非职业科学家（如科学爱好者等）组织或参与的科学研究活动，业余天文学(citizen astronomy)是公众科学的一个经典分支。受益于科技水平的提升，当代天文爱好者有力地补充了职业天文学家无力或无暇顾及的领域，如时域天文学观测、大数据的人工分析、数据挖掘等。近年来，我国业余天文学发展迅速，我国爱好者在新天体的搜寻和发现上成绩显著，但与欧美国家的业余天文相比，我国爱好者的兴趣点比较单一，对长期监测、数据挖掘等项目的参与度低。调查发现，我国青少年在业余天文学家群体中占有很大的比重，"对天文感兴趣"、"学习天文知识"、"获得乐趣"和"认识朋友"是我国爱好者参与业余天文学研究的主要动机，这两点与欧美国家业余天文学家有着明显不同。随着我国一系列大科学设备的建成运行，公众科学和业余天文学的潜力需要得到重视。应对公众和爱好者进行积极引导，使他们在学习知识和获得乐趣的同时，能为科学研究作出更大贡献。

**关键词**：天文学的社会属性
**中图分类号**：TN216  **文献标识码**：A  **文章编号**：0000-0000-(2017)0-0000-00

业余科学家指从事科学研究、但并不以此为职业的人[1].。因为业余科学家的工作往往来自于强烈的兴趣驱动，他们应被视作科学爱好者群体的一部分：科学爱好者泛指对科学感兴趣的人，而业余科学家则特指有针对性地从事科学研究的爱好者。出于语言习惯的考虑，以下我们不作细分，将任何从事科学研究的业余科学家统称为"爱好者"，但我们应注意业余科学家相对于一般爱好者的独特性。同时我们还要注意到业余科学家和所谓的"民间科学家"(crank)的明显区别："民间科学家"多不了解也不接受科学共同体的研究模式，而业余科学家仍然学习和使用科学方法来系统性地研究问题。

"业余"天文学家与"职业"天文学家的概念划分最早出现于 18 世纪末的欧洲[2]。职业天文学家供职于国家或教育机构的天文台，而业余天文学家则一般自建天文台来追求自己的兴趣。作出重要贡献的业余天文学家也可能被国家或教育机构雇佣而转为"职业"，比如发现天王星的赫歇尔(William Herschel)和从事彗星搜索的庞斯(Jean-Louis Pons)。类似的划分在 19 世纪中叶也出现在北美和澳洲[3][4].，业余天文学家亦开始做出更多贡献，如拉瑟福德(Lewis Morris Rutherfurd)和罗伯茨(Isaac Roberts)等人在天文摄影术的探索。据估计，目前北美的业余天文学家群体约有 500 人左右，占北美天文爱好者总数的 0.1%，数量上也仅为职业天文学家人数的 1/10[5]。在双语文献中，我们还应注意中文与英文在语义上的细微差比："业余天文学家"对应的英文不是 amateur astronomer（语义更接近"天文爱好者"），而应是 citizen astronomer（直译为"公众天文学家"）。以下我们遵循现有的用语习惯，将 citizen astronomy(-er)翻译成业余天文学（家）。

我国古代的"天学"原为帝王之术，禁止百姓私习，这一情况在封建帝制结束之后有所改观。随着我国人民生活水平的提高，我国的天文教育也快速发展，开始有天文爱好者涉足研究领域，甚至发表同行评议的论文。随着互联网和大数据时代的到来，天文学也成为一门数据密

集型学科，业余（公众）天文学以及公众科学的发展日新月异。Marshall 等人对业余天文学的发展和现状进行了详细评述[6]，但主要研究对象是欧美国家。我国的业余天文学起步晚，发展快，存在一定的特殊性。因此，本文将尝试对我国业余天文学的现状进行研究和评述。

# 1. 业余观测

天文学最早起源于对夜空的观测，目前观测仍然是天文爱好者的主要活动。职业天文学家与爱好者的硬件水平原本差异不大，但随着科技的进步和各国对基础科学研究的重视，这一差异在 20 世纪变得愈发明显，职业天文学研究早已不局限于可见光波段的研究，但大多数爱好者仍然专注于可见光天文学。不过，尽管可见光天文学的发展已经趋进饱和，时域天文学(time domain astronomy)的兴起以及信息科技的飞速发展仍然为爱好者提供了相当的空间。如今，爱好者级别的望远镜和 CCD 相机获得的相片已经可以媲美 30-40 年前大望远镜和昂贵的照相底片才能取得的相片，便携式光谱仪等也开始步入商业市场，拓宽了爱好者的研究范围。目前，天文爱好者在暂现事件(transients)和太阳系小天体的发现和观测上仍发挥着相当作用。我国爱好者虽起步较晚，但即使在 70-80 年代仍有所成绩，如我国段元星等人独立发现 1975 年天鹅座新星(V1500 Cyg)，周兴明独立发现 C/1983 H1 等 12 颗彗星，欧阳天晶对流星活动的无线电观测等。进入 21 世纪，我国经济的飞速发展以及互联网时代的到来迅速缩小了我国与国外爱好者的差距，体现在硬件、软件配备以及活动思路上。

## 1.1 业余程控天文台的兴起

现代文明带来的光污染及大气污染，使得天文台站不得不向越来越偏远的地区迁移。与此同时，计算机的小型化以及网络技术的飞速发展使得远程观测成为可能。早在 1975 年，Colgate 等人已实现用微波网络远程控制一架 75 厘米的望远镜[7]。能够自主执行观测并适应环境变化（诸如起云、下雨等），同时不依赖人工干预的"程控自主天文台"(Robotic Autonomous Observatory, RAO)在 1984 年问世。RAO 技术被大量应用于目标明确的观测项目，比如超新星巡天、伽马暴余晖观测等。Castro-Tirado 等[8]和崔辰州等[9]详细介绍了 RAO 的历史与现状。

远程观测以及 RAO 技术在 1990 年代末开始在天文爱好者中使用。Bob Denny 开发的 Astronomy Common Object Model(ASCOM)协议及相关软件解决方案 [1]，在业余天文台站中使用广泛，甚至部分专业天文台站也有使用。相较于职业天文学家而言，爱好者的工作模式具有较大的随意性，人工干预不仅不是技术壁垒，反而是观测活动的组成部分之一。因此虽然许多业余天文台的程序化程度已经很高，但能达到"程控自主"阶段的天文台仍然较少，主要是用于教育或商业目的的网络望远镜，比如布拉特福德望远镜 [2]、iTelescope 网络 [3]等。

建立于 2007 年、台址位于中科院新疆天文台南山基地的星明天文台是我国第一座稳定运行的业余程控天文台。星明天文台由乌鲁木齐的高兴主持建设并运行，主要承担爱好者主导的超新星、彗星巡天项目，其获取的数据在网络上公开，供全国各地数百位爱好者分析研究。星明天文台也与中国虚拟天文台合作开展公众超新星搜寻项目 [4]。除此以外，该台也利用地理区位优势加入职业天文学家主导的时域天文研究[10][11]。从国际天文联合会(International Astronomical

---

[1] http://acpx.dc3.com/
[2] http://www.telescope.org/
[3] http://www.itelescope.net/
[4] http://psp.china-vo.org/

Union, IAU)小行星中心天文台数据库来看，目前国内活跃的业余观测台站，还有苏州绿野天文台等4处（图1）；此外还有至少十余座以天文摄影为主的远程天文台，如西藏羊八井星河科研社天文台、新疆南山若岩天文台等。一份不完整的名单列于表1。

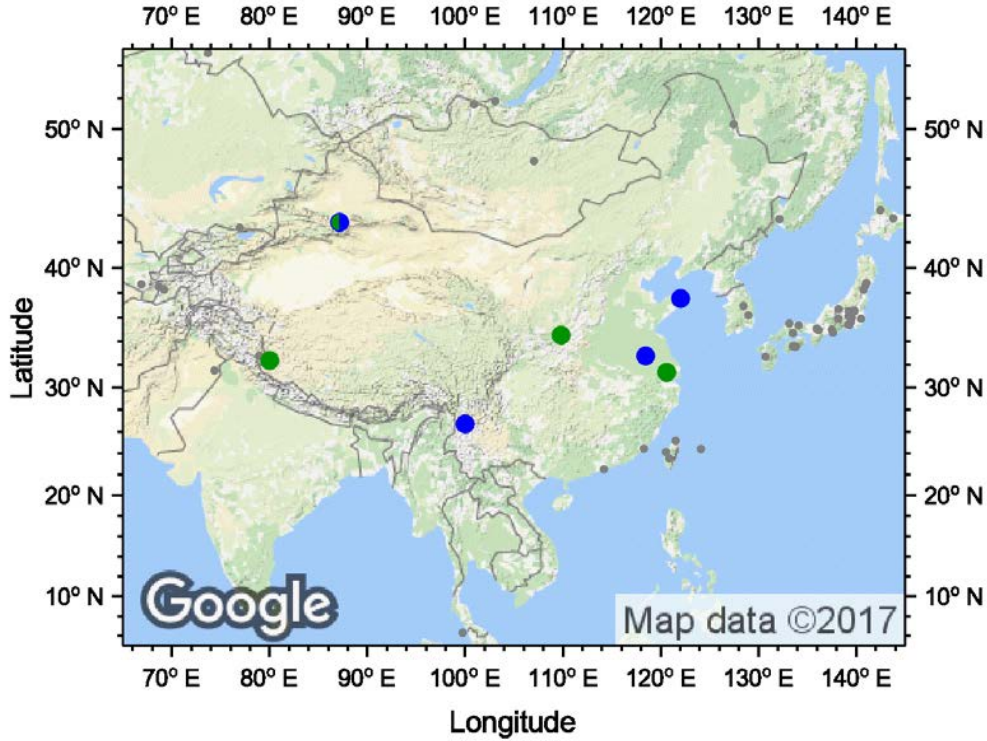

图 1. 在过去 5 年内(2012-2017 年)有向国际小行星中心上报数据的中国大陆台站。蓝色圆点为专业台站，绿色圆点为业余台站，蓝绿各半的圆点表示此地既有专业也有业余台站。港澳台地区及国外台站以灰色圆点表示。

Figure 1. Active astronomical observatories in mainland China that have submitted observations to the IAU Minor Planet Center in the past 5 years (2012–2017). Blue data points represent professional observatories; green data points represent citizen-built observatories. Mixed color points represent observatory sites that are occupied by both professional and citizen-own facilities. Observatories in Hong Kong, Macau, Taiwan and other countries are shown as gray dots.

表 1. 中国大陆部分业余天文台站列表。
Table 1. An incomplete list of active citizen-built observatories in mainland China.

| 站点名称 [5] | 地点 | 建设年份 | IAU 编号 | 设备 |
|---|---|---|---|---|
| 星明天文台 | 新疆乌鲁木齐 | 2007 | C42 | 50 厘米反射、25 厘米折反射等 |
| 若岩天文台 | 新疆乌鲁木齐 | 2010 | | 25 厘米反射、10 厘米折射等 |
| 星河科研社天文台 | 西藏羊八井 | 2011 | | 50 厘米反射 |
| （敷山） | 陕西少华山 | 2012 | G26 | 26 厘米折反射 |
| 北冕天文台 | 西藏阿里 | 2013 | N55 | 30 厘米反射 |
| 绿野天文台 | 江苏苏州 | 2013 | P34 | 30 厘米反射、25 厘米反射 |

---

[5] 部分台站的名字无法查得，因此以括号注明台站所在地。

| | | | | |
|---|---|---|---|---|
| ULTRA 天文台 | 江苏苏州 | 2013 | P36 | 25 厘米反射 |
| 碧林天文台 | 河北石家庄 | 2014 | | 36 厘米反射、20 厘米折反射 |
| （吴川） | 广东湛江 | 2014 | | 15 厘米反射 |
| （阎良） | 陕西西安 | 2015 | | 13 厘米折射 |
| 紫垣天文台 | 湖北红安 | 2015 | | 16 厘米反射 |
| 北斗天文台 | 云南武定 | 2016 | | 20 厘米反射等 |
| 北京 KYP 天文台 | 北京 | 2016 | | 11 厘米折射 |
| 日月星尘天文台 | 江西婺源 | 2016 | | 10 厘米折射 |
| 松涛天文台 | 河南郑州 | 2016 | | 13 厘米折射 |
| 星光之友天文台 | 湖北郧西 | 2016 | | 25 厘米反射、20 厘米反射等 |
| 玉海天文台 | 广西博白 | 2016 | | 20 厘米反射 |
| 双子天文家园 | 云南丽江 | 2017 | | 16 厘米反射、11 厘米折射等 |

  这些业余程控天文台的台站建筑多为爱好者自行设计建造，大多数位于郊区或野外的自有地，但也有少数与科研机构合作，建造在科研机构所有的地块内（如北冕天文台、星明天文台、星河科研社天文台）。所用的设备及配套软硬件（望远镜、相机、观测设备控制等）绝大多数来自成熟的商业产品。近 2/3 的天文台建成于过去 3 年，可见我国业余程控天文台的数量呈快速增加的趋势。但只有不到 1/3 的台站获得了国际天文联合会编号。实际上，大多数程控天文台主要用于天文摄影，这些台站所有者的主要活动多以摄影作品的形式发表在网络媒介上（如天文论坛、微博、微信公众号等），几乎没有涉及科学研究工作，像星明天文台这样由爱好者自发组织进行科学研究的例子十分少见。

  值得注意的是，国内职业天文学家也开始注意到爱好者的工作，并和他们展开合作，这表明我国天文界的专业-业余合作(Pro-Am Collaboration)已经起步。典型的例子有星明天文台协助紫金山天文台确认近地小行星 2017 BM$_3$[12]以及清华大学的王晓峰课题组帮助认证星明天文台发现的超新星[13]等。

## 1.2 暂现源及太阳系小天体的观测

  暂现源指可见时间短暂的天文事件，如超新星爆发、新星增亮、太阳系小天体（小行星、彗星等）的回归等等。由于多数暂现源的出现无法提前预知，专业大型设备很难充分发挥其长处，因此为爱好者留下一定的活动空间。爱好者在新天体的发现上贡献尤多：直到不久前，业余爱好者发现的超新星占每年超新星发现总量的 10% 以上[6]；近年一些比较明亮或独特的新彗星也是爱好者发现的，比如 C/2011 W3 (Lovejoy)、332P/Ikeya-Murakami 等[14][15]。此外，由于爱好者的分布远较职业天文学家的分布更广，他们往往能对新事件或新天体进行快速跟踪观测，为日后科学研究提供重要数据，如 2008 年撞击地球的小行星 2008 TC$_3$、2015 年爆发的彗星 15P/Finlay 等[16][17]。

  即使在上世纪 70--80 年代，我国也有爱好者活跃于新天体的搜寻，如段元星、周兴明等，但由于技术和通信上的限制，所作出的发现大多晚于国外天文学家及爱好者（不过，尽管没有得到命名权，周兴明独立发现 C/1990 N1 (Tsuchiya-Kiuchi)和 122P/de Vico 仍然为国际天文联合会所认可）。我国爱好者首次真正意义上的领先发现，应属河南张大庆在 2002 年发现的池谷-张彗星(153P/Ikeya-Zhang)。在此之后，我国爱好者又发现了 3 颗彗星[C/2008 C1 (Chen-Gao), 325P/Yang-Gao, C/2015 F5 (SWAN-Xingming)]、至少 32 颗超新星、6 颗新星以及将近 100 颗小行星，几乎全部由星明天文台发现。除了新天体搜寻及发现以外，对已知天体的跟踪观测也是爱好者经常进行的课题，如星明天文台加入全球彗星联测网[18]，绿野天文台的刘君达对(2729)

Urumqi 等多颗小行星的自转周期测量[19][20][21]等。此外，小行星掩星也很受爱好者关注。小行星掩星是一种特殊的暂现现象，只有掩星带内才能观测到；而爱好者分布广泛、机动性高，在掩星事件的监测上具有很大优势。在这方面作出工作的有张学军、陈栋华等等 [6]。

## 1.3 流星观测

流星本质上是一种大气现象，但其成因与天文学研究对象直接相关，因此一般被纳入天文学的研究范畴。由于流星可在任何时间、任何地点出现，且单颗流星只能流星出现点下方不大的一块区域内看见，难于用常规的天文学研究手段大量获取数据，因此爱好者在流星研究上发挥着很大作用。

目视流星观测是一项历史悠久又简单易行的活动。历史上，目视观测法也是最常用和有效的流星观测方法之一，直到近年才逐渐被流星摄像监测网取代。国际流星组织(International Meteor Organization, IMO)是国际流星研究的专业组织，汇集自 1982 年起世界各地的目视流星观测报告并作分析研究。根据已发表的文献来看，我国现代的目视流星观测最早开始于 20 世纪 80 年代对宝瓶座 η 流星雨的监测[22]，参与 IMO 联测的最早者则是武汉的欧阳天晶（1989 年）。随着我国天文爱好者群体的壮大，我国参与 IMO 联测的人数也迅速增加。从近年英仙座流星雨的国际联测活动来看，有 10-20 的报告都来自中国内地（图 2；有趣的是，近年唯一一次无中国爱好者参与的英仙座流星雨联测活动为 2008 年，原因可能是英仙座流星雨活动时间正值北京奥运）。但尽管参与人数有明显的上升趋势，我国观测者的总观测时长仍然很短，仅占全球总观测时长的 1% 不到。即使是观测数据贡献最多者，每年也仅贡献 10 小时左右的数据，低于 IMO 的平均值（30 小时/人），说明我国能坚持进行目视流星观测的爱好者仍然很少。

---

[6] 见陈栋华，"小行星掩星观测攻略"，《天文爱好者》2005 年第 4 期；张学军，"联测 10199 号小行星掩星"，《天文爱好者》2015 年第 6 期。

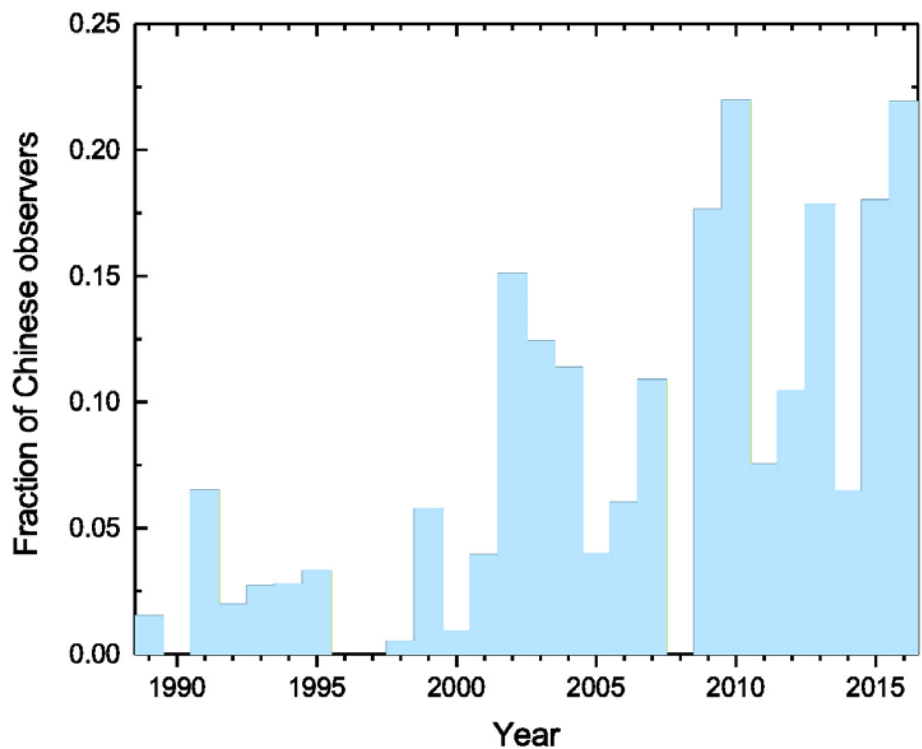

图 2. 中国内地英仙座流星雨观测报告所占比例随时间的变化。数据来自国际流星组织英仙座流星雨联测分析报告。
Figure 2. Fraction of Perseid observing reports made in mainland China versus time. Data obtained from the annual analysis of Perseid meteor shower made by the International Meteor Organization.

流星摄像监测网(video meteor network)利用摄像设备取代人眼记录流星，可以大大节省人力支出，还可以尽量排除观测者主观因素的影响。近年来，随着摄像设备的小型化和网络化，流星摄像监测网得到很大发展，其中相当大的比例仍然由业余天文学家建设和运行。后期数据处理多依赖于商业化解决方案，如SonotaCo的UFO系列软件[23]。我国第一个固定运行的流星监测网是北京天文馆主持建设的火流星监测网，自2010年开始运行；第一个由爱好者建设的流星监测网是广东流星监测网[7]，自2012年开始运行。至今为止，各地爱好者还在山东青岛、新疆乌鲁木齐和西藏阿里一带建成流星监测网，简要情况参见表2。除了流星的摄像监测以外，也有部分爱好者尝试对流星进行光谱观测，如程思浩和程思淼的工作[24]。

表2. 中国内地流星摄像监测网简表。
Table 2. A brief summary of the video meteor networks in mainland China.

| 地点 | 建设年份 | 站点数量 |
| --- | --- | --- |
| 北京 | 2010 | 6 |
| 广东广州[8] | 2012 | 3 |
| 山东青岛[9] | 2015 | 3 |

---

[7] http://www.theskylab.org/

[8] http://www.theskylab.org/

[9] http://tqw.lamost.org/zidong.htm

| | | |
|---|---|---|
| 新疆乌鲁木齐 | 2016 | 7 |
| 西藏阿里 | 2017 | 2 |

流星产生的余迹能反射电磁波,因此可以在无线电波段观测流星。无线电流星观测分为主动和被动两种,观测者主动发射无线电脉冲并利用反射的信号观测流星,称为主动观测;观测者接收远处发射台(如远方的调频广播台等)的信号,称为被动观测。被动观测因为所需仪器简单,更为观测者常用。武汉的欧阳天晶在1990年代起即坚持进行无线电流星观测,并参与了国际无线电流星联测项目[25]。上文提及的广东流星监测网也在进行被动无线电观测。主动无线电观测一般利用雷达进行。雷达成本较高,且涉及政策管控问题,很少有爱好者使用。但无线电爱好者常用的业余无线电频段也可以进行主动无线电观测。大庆的张学军在1998年狮子座流星雨期间曾进行过这类试验[10]。

# 2. 数据挖掘

天文学早已迈入大数据时代,各种时域天文学巡天项目如卡塔琳娜巡天(Catalina Sky Survey)、泛星计划(Pan-STARRS)和大口径综合巡天望远镜(Large Synoptic Survey Telescope)等,已经或即将产生千兆字节量级的数据[26]。这些海量的数据不仅帮助职业天文学家以更高的时间分辨率研究他们感兴趣的目标,还为爱好者提供了广阔的探索空间。爱好者参与大数据挖掘主要有两种方式:第一种是爱好者自发进行的探索,比如在巡天项目发布的图片中搜寻新天体;另一种是职业天文学家设计好课题,请爱好者协助分析获取的数据,比如星系动物园(Galaxy Zoo)、(系外)行星猎手(Planet Hunters)等。

## 2.1 爱好者自发进行的研究

爱好者对数据的自发探索几乎完全以发现新天体为主要目标。1990年代后期,互联网在我国开始普及,极大地降低了我国爱好者获取数据资料的技术门槛。第一个利用公开数据搜寻并成功发现新天体的是新疆的周兴明,他在2000年发现彗星C/2000 X4 (SOHO)。SOHO(太阳和太阳风层探测器,Solar and Heliospheric Observatory)卫星是美国航空航天局与欧洲航天局联合发射的一颗太阳观测卫星,其主要目的是研究太阳,但意外发现其日冕仪可以用来观测掠日彗星。自1996年SOHO课题组科学家S. Stezelberger发现首颗SOHO彗星之后,已有100多位世界各地的研究人员(绝大多数为业余爱好者)成功从SOHO观测的数据中发现彗星。自周兴明发现首颗SOHO彗星后,我国SOHO彗星搜寻者的群体迅速壮大。至2013年8月为止,我国共有25位爱好者成功发现SOHO彗星[11],占所有SOHO彗星发现者总数的1/4,其中陕西周波以287颗彗星的总发现数居全球彗星猎手排行榜之首。

除SOHO数据库以外,NEAT数据库亦常为爱好者用于发现新天体。NEAT为"近地小行星追踪"(Near-Earth Asteroid Tracking)的缩写,项目使用帕洛玛天文台1.2米施密特望远镜以及夏威夷哈里亚克拉1.2米反射镜搜寻近地小行星,至2010年止共拍摄了69万张图片,所有图片可在SkyMorph网站获取[12]。自2001年开始,开始有爱好者利用这一数据库搜寻被遗漏的小行星。我国爱好者自2005年发现首颗NEAT小行星以来,至2014年7月为止,累计已有7位爱好者发现655颗小行星[13]。除了极早期的爱好者发现以外,NEAT数据库中发现的小行星,发现权及命名权均

---

[10] 见寇文,"流星的业余无线电观测",《星光快讯》2002年第6期。

[11] 见孙霈源,http://cometobserver.lamost.org/search/comet/sohorank.pdf

[12] https://skyview.gsfc.nasa.gov/skymorph/skymorph.html

[13] http://www.skaw.sk/discoverers-list-skymorph-archive.htm

归属NEAT课题组。

## 2.2 职业天文学家主导的研究

职业天文学家主导的业余天文学研究可大致分为两类：目视分析，指公众协助职业天文学家，按照指定的方式对大量数据进行人工分析；志愿计算，指公众利用空闲的计算资源，协助职业天文学家进行大数据的处理。

公众科学目视分析项目的产生直接源自于现代天文巡天产生的海量数据：形态复杂的目标难以完全依赖计算机进行研究，而研究人员的人力资源又有限。第一个实现公众科学目视分析的Stardust@home[14]，志愿者们需要人工检查星尘号探测器拍摄的接近3000万张照片以寻找星际尘埃留下的痕迹[27]。继承了Stardust@home的星系动物园项目[15]则需要志愿者对数十万个星系进行人工分类。星系动物园项目大获成功，在起基础上诞生了横跨天文学、生态学、生物医学等多个学科的Zooniverse平台，公众科学家可以在上面寻找自己感兴趣的课题加入。这些项目和平台主要使用英语进行交流。尽管有志愿者提供了内容简介和分析指引的中文译本，但我国爱好者的参与程度仍然不高。以星系动物园为例，Raddick等在2013年进行的调查表明，来自中国的参与者只占被调查用户总数的0.8%，而来自英语国家（美国、英国、加拿大、澳大利亚等）的用户所占比例则为70%左右[28]。相比之下，我国爱好者参与程度较高的目视分析项目，仍然集中在以发现新天体为主要目标的项目，如亚利桑那大学Spacewatch课题组发起的快速移动天体(Fast Moving Object)搜索项目[29]，至2005年该项目结束止，我国爱好者占志愿者总数的20%左右，共发现了6颗近地小行星；依托于国家天文台的中国动手天文(Hands-On Universe, HOU)在2008-2010年前后也组织北京等地的学生参加国际小行星搜寻活动，共发现7颗小行星[16]。这说明语言障碍不是阻碍我国爱好者参与国际项目的主要因素。

志愿计算只需要利用空闲的计算资源，对志愿者来说参与成本更低。第一个业余天文领域的志愿计算项目是搜寻外星信号的SETI@home，在此基础上发展出伯克利开放式网络计算平台(Berkeley Open Infrastructure for Network Computing, BOINC)，吸纳了许多不同领域的志愿计算项目。在天文领域的项目中，SETI@home仍然是其中最热门的项目，除此以外还有搜寻脉冲星引力波的Einstein@home，对银河系进行建模的MilkyWay@home，以及通过光变曲线分析小行星形状的Asteroid@home等等。尽管有专门的中文站点推广BOINC（中国分布式计算总站[17]），中国科学院高能物理研究所还推出了单独的计算项目CAS@home，专门用于支持中国科学家主导的项目，但从BOINC网站上的统计来看，我国的贡献仍与欧美国家甚至日本有明显差距（中国大陆的月平均得分排在252个国家和地区的第17位，台湾地区排第16位），且用户的平均贡献也较低（排在第187位）。这再次说明语言障碍和政策导向不是我国爱好者参与意愿较低的主要因素。

除以上介绍的方向之外，还有一些爱好者在其他领域有所建树，如研究脉冲星磁场振荡模型的梁助兴[30]、考证小行星命名历史的林景明[18]、研究天文学史并著书的潘鼐[19]等。但总得来说，业余观测和数据挖掘仍然是业余天文学家主要涉及的领域。

---

[14] http://stardustathome.ssl.berkeley.edu/

[15] https://www.galaxyzoo.org/

[16] 见郭红锋，"第四期国际小行星搜寻活动捷报频传"，《中国科技教育》2010 年第 1 期。

[17] http://www.equn.com/

[18] 见林景明，"寻根究底追 Juewa——139 号小行星姓甚名谁"，《天文爱好者》2017 年第 7 期。

[19] 如潘鼐，《中国恒星观测史》，学林出版社。

# 3. 讨论

## 3.1 业余天文学家的动机

是什么驱动着业余天文学家们进行他们的研究呢？Price和Paxson、Raddick等人分析了欧美国家的爱好者参与数据挖掘项目的主要动机[28][31][32]并指出，对天文（或者所研究领域）的兴趣、对科学研究的热情以及对宇宙壮丽的欣赏都是主要因素，而在研究过程中结识朋友、学习新知识和得到乐趣则相对不那么重要。然而，针对中国业余天文学家的调查仍然缺乏。因此，我们设计了一份调查问卷，发放给社交软件QQ上"星明天文台巡天群"的参与者。调查内容包括参与者所在省市、年龄阶段、性别、最高学历、职业，是否职业天文工作者，以及主要参与动机。参与动机共分为8项，动机的选择参考了Raddick等人[28]的调查结果，分别是：

1. 为科学研究做贡献
2. 学习天文学有关知识
3. 发现新天体
4. 认识志同道合的朋友
5. 教授知识和帮助他人
6. 欣赏宇宙之美
7. 获得乐趣
8. 对天文感兴趣

我们回收了105份有效问卷，在排除职业天文工作者（包括天文专业的学生）后，共有有效问卷97份。这是否能有效代表中国的业余天文学家群体呢？根据社交网站相关话题的热度以及天文论坛的用户数来看（如"知乎"关注"天文学"话题的用户数，百度"天文吧"的用户数等），关注天文类话题的用户约为10--100万量级；每日活跃用户（如"牧夫"天文论坛、"天之文"论坛）以及针对性较高的讨论区的用户数（如"知乎"关注"天体物理学"话题的用户数，百度"天文爱好者吧"的用户数等）大约在1000--10万之间，由此我们估计国内天文爱好者的数量应在1--10万这个量级。按照北美地区天文爱好者与业余天文学家的比例[5]，国内的业余天文学家群体应在10--100人左右。同时北美地区职业天文学家与业余天文学家之比约为10:1[5]。考虑到中国天文学会目前的会员数为2000左右[20]，国内业余天文学家的数量应在200上下。我们可以注意到这几个数字在数量级上相互吻合，因此可以认为我们的调查非常有效地覆盖了中国的业余天文学家群体。

参与者的统计情况以及和星系动物园统计结果的比较参见图3及图4。从参与者的基本构成来看，我们观察到以下几点：

1. 过半的参与者来自经济发达省份。
2. 参与者大多为男性(92%)，这与星系动物园的调查结果相近(82%)，但男女比例更为悬殊。
3. 过半的参与者为学生(60%)，其次是专业人员(12%)、普通职员(8%)、企业管理者(5%)和自由工作者(4%)，来自这五类职业的参与者占总参与者的90%。学生和专业人员所占比例大大高于我国互联网用户中的对应比例(25%，5%)[21]。
4. 年龄构成也印证了职业构成的结果：69%的参与者不到25岁，显示出学生所占比例很高。这与星系动物园参与者的年龄构成（参与者集中在25-55岁年龄段）完全不同，15-24岁

---

[20] 参见《科技导报》2011年第21期。
[21] 参见中国互联网络信息中心，《第39次中国互联网络发展状况统计报告》，
http://www.cnnic.cn/hlwfzyj/hlwxzbg/hlwtjbg/201701/P020170123364672657408.pdf

年龄段的参与者也明显高于对应年龄段的互联网用户所占比例，显示出我国青少年一代有很高的参与热情。

5. 参与的学生包括大、中学生（可能还有部分小学生，但调查问卷未作细分），但以高中生为主。在已迈入工作岗位的参与者中，持有本科或专科学历的参与者占多数，但也有少量参与者为高中学历。有意思的是，参与者中没有已获得研究生学历的人，但有在读研究生（占总调查群体的3%），这可能与我国青年一代受教育程度迅速提升有关。这与星系动物园的调查结果有相似之处，即已毕业的参与者以高学历者为主，但星系动物园的参与者普遍有研究生学历(54%)。

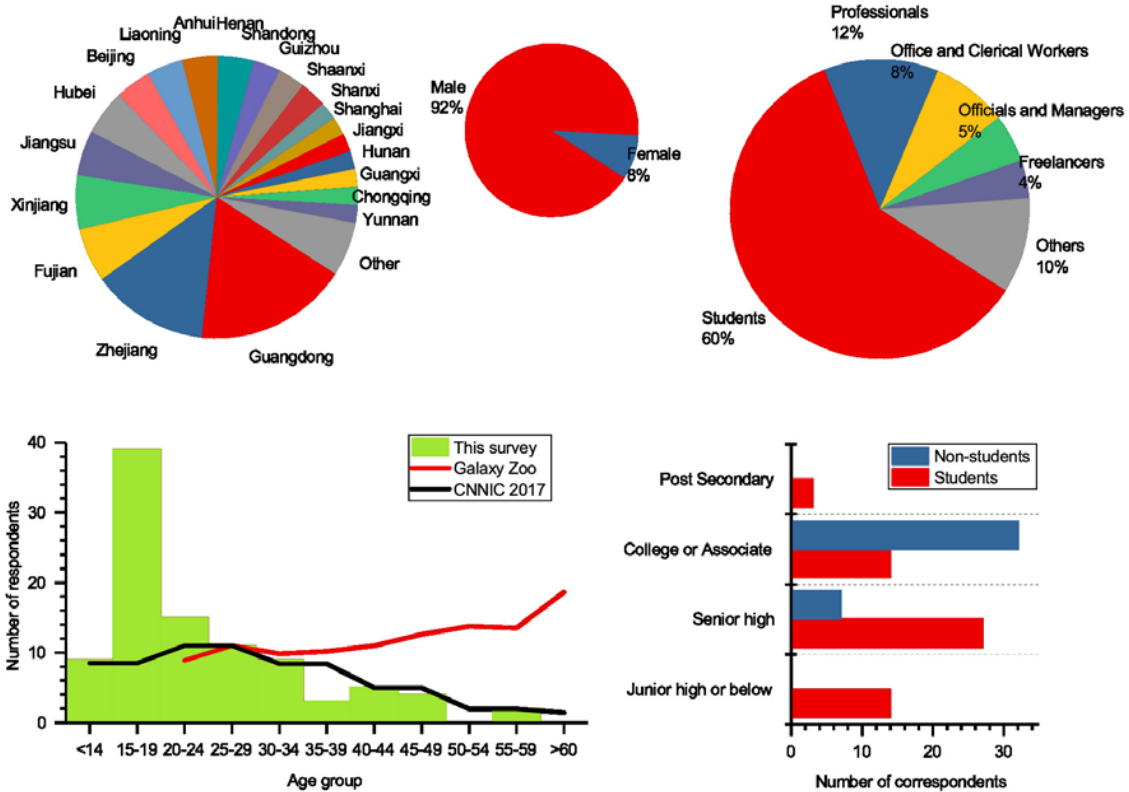

图3. "星明天文台巡天群"调查参与者的基本情况，第一行从左到右分别为省市、性别、职业分布，第二行从左到右分别为年龄和学历构成。星系动物园的数据来自Raddick等人的工作[28]；中国互联网用户的年龄分布来自2017年1月发布的第39次《中国互联网络发展状况统计报告》。中国互联网用户年龄分布的分组较另外两组数据更宽，为10年一组。为了便于比较，星系动物园和中国互联网用户的数据均乘以一个系数，使得25--29岁年龄段的数值与本调查的数值相同。
Figure 3. Demographics of the participants that take the questionnaire survey of the Xingming Observatory Survey Group. The upper panel is the geographic distribution (left), gender distribution (center) and occupation distribution (right). The lower panel is the age and education distribution (left to right). Age distribution from Galaxy Zoo is extracted from the work of Raddick et al. [28]; age distribution of the Internet users in the People's Republic of China is extracted from the 2017 Annual Report from the China Internet Network Information Center, which is binned differently (10 years per group) compared to the other two sets of data. For clarify, the data from the Galaxy Zoo and Chinese Internet users have been scaled such that the data values of the 25--29 years old group equal to that of the questionnaire survey.

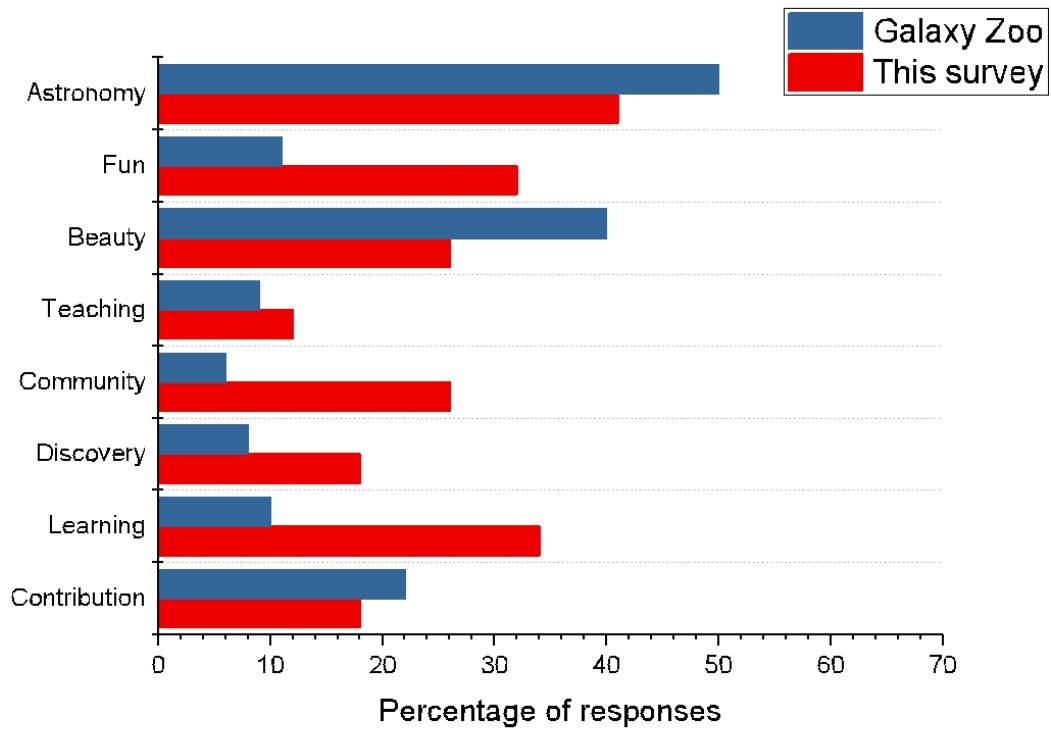

图4. "星明天文台巡天群"调查参与者的参与动机与星系动物园调查结果[32]的比较。本调查的动机选取与星系动物园调查的动机[32]选取大致相同，只是将"教授知识"(Teaching)和"帮助他人"(Helping)，"欣赏宇宙之美"(Beauty)和"欣赏宇宙的广博"(Vastness)，以及"对天文感兴趣"(Astronomy)和"对科学感兴趣"(Science)各合并为一项，同时去掉了"星系动物园"(Zoo)和"其他"(Other)两项。

Figure 4. Motivation of the participants that take the questionnaire survey of the Xingming Observatory Survey Group versus the result of the Galaxy Zoo survey [32]. The choice of the motivation elements is largely identical to the Galaxy Zoo survey [32], except for merging ``Teaching'' and ``Helping'', ``Beauty'' and ``Vastness'', ``Astronomy'' and ``Science'' into three single elements, and removing the elements of ``Zoo'' and ``Other''.

　　从参与动机的构成来看，最主要的参与动机分别是对天文的兴趣(共有41%的受访者认为这是最重要的参与动机)，最不重要的参与动机分别是教授知识和帮助他人(12%的受访者)。对天文的兴趣是我国爱好者参与星明天文台巡天项目的主要动机，教授知识和帮助他人是次要动机，这两点和星系动物园调查的结果相似；但本调查的参与者认为学习知识、获得乐趣和认识朋友也是关键因素，这与星系动物园的参与者不同。

　　值得注意的是，虽然前文提及我国爱好者热衷于发现新天体并有显著成绩，但从调查结果来看，发现新天体不是爱好者参与星明天文台巡天的最主要动机,这一对结果是否相互矛盾呢？我们认为，这说明我国爱好者的研究兴趣具有一定的可塑性，并不局限于新天体发现。近年来我国爱好者屡屡发现新天体，在国内天文爱好者群体内引起了很大反响；而新天体搜索上手简单，周期短、见效快，容易给爱好者带来乐趣，大量的参与者形成了自己的小社区，满足了爱好者"学习天文知识"、"获得乐趣"和"认识朋友"的需求。长期监测、数据分析等项目周期长，参与门槛高，难于满足爱好者的需求，因此参与度低。但若加以合适引导，爱好者对这些项目的参与度是有提升空间的。

此外，虽然语言障碍不是阻碍我国爱好者参与国际项目的主要因素，但客观上也增加了我国爱好者与国外爱好者交流的难度。国外爱好者办有同行评议的刊物供业余科学家们发表成果及交流，来自我国的文章也很少见（表3）。这些现象固然与爱好者的个人兴趣有关，但也反映出我国爱好者缺乏有效的引导和激励，导致成果出口集中于短、平、快的项目。

表3. 部分供爱好者进行交流的同行评议刊物。
Table 3. A partial list of peer-review journals for citizen astronomers.

| 期刊 | 创刊年份 | 2016年中国作者发表比例[22] |
| --- | --- | --- |
| Journal of the American Association of Variable Star Observers<br>美国变星观测者协会会刊 (JAAVSO) | 1972 | 0/78 |
| Minor Planet Bulletin<br>小行星快讯 (MPBu) | 1974 | 5/128 |
| Journal of Occultation Astronomy<br>掩星天文期刊(JOA)及<br>Occultation Newsletter掩星快讯(ON) | 1974 | 0/10 |
| WGN, Journal of the International Meteor Organization<br>国际流星组织会刊 (WGN) | 1987 | 0/43 |
| Journal of Double Star Observations<br>双星观测期刊 (JDSO) | 2005 | 0/84 |

## 3.2 中国业余天文学的未来

对于热衷观测的业余天文学家来说，大数据天文学和时域天文学的发展正快速蚕食他们的空间，超新星、彗星等新天体的发现已经基本被职业天文学家所垄断，新发现的天体越来越暗弱，爱好者难于协助确认。下一代快速巡天项目，如兹维基巡天设施(Zwicky Transient Facility)和LSST的上线，将进一步压缩业余爱好者的空间。此外随着爱好者采集的数据量不断增加，他们自己也会面临大数据的处理问题，比如搜寻新天体的爱好者往往需要编写或购买商用程序来处理更大量级的数据。但应该注意的是，下一代快速巡天项目多位于西半球，而东半球的观测设施仍然相对缺乏，我国爱好者的地理位置优势使得他们仍然在一定程度上能与职业天文学家实现互补。因此，对于变化速度很快的暂现事件（如明亮的超新星事件、彗星爆发等），我国爱好者仍然具有很大潜力。此外，对职业天文学家较少或较难涉及的"冷门"领域，爱好者也有很大的发展空间，如流星观测、陨石目击及搜寻、大行星监测等。

在大数据处理和分布式计算方面，我国的网民数量位居世界第一，智能手机的普及率也居世界前列，公众参与大数据科学有着很大潜力。中国虚拟天文台和星明天文台合作运行的公众超新星搜寻项目以及CAS@home项目都是很好的尝试。我国已建成五百米口径球面射电望远镜，加入平方千米阵，准备建设12米大视场光学红外望远镜等新的光学巡天设备。这些大科学装置将会产生大量数据。依靠大数据处理和分布式计算可以进一步对这些数据进行挖掘，提升其科学产出。因此，这些大科学装置的建成应被视为我国公众科学发展的契机。

我们的调查发现，青少年占据着我国天文爱好者群体中极大的比重，我国爱好者对"寓教于乐"式的学习也有着很大兴趣。尽管我国业余天文学家此前的贡献主要集中在新天体的发现上，但获得新发现并不是他们参与天文研究的主要动机，因此他们的研究兴趣具有很大的可塑性。在这方面，欧美日等国的经验值得我们参考和借鉴，比如：

---

[22] 2016年全年，中国爱好者以第一作者身份发表的文章数与总文章数之比。

- 提供专供爱好者发表学术小论文的刊物（或者在已有刊物上提供版面或发行增刊），引入同行评议、英文摘要等学术共同体通行的制度，引导爱好者将他们的工作系统化，同时鼓励国内不同地域和国内外爱好者之间的交流。
- 鼓励高阶爱好者建立定期交流机制，能对共同感兴趣的问题进行严肃的讨论，同时鼓励有兴趣和能力的爱好者参与专业的学术会议。
- 创造机会让爱好者了解科研领域的新进展和新技术，如数据自动化处理、机器学习、大数据管理等。

总之，如何让我国公众和爱好者认识到科学除了新发现以外的其他研究方向，提升他们对科学研究的总体热情，是一个值得思考的问题。

## 4. 结语

我国的业余天文学在最近一二十年来取得了很大发展，我国天文爱好者在新天体的搜寻和发现上成绩尤其显著。但与欧美国家相比，我国业余天文学在广度和深度上还有不小的差距，公众和爱好者的兴趣单一，职业科学家、爱好者和公众的互动仍然较少。随着经济水平和受教育程度的不断提升，我国公众对自然科学的兴趣也在不断增加，天文爱好者群体也在迅速发展壮大，我国的业余天文学有着巨大的潜力。实现这一潜力有助于促进我国科研和教育实力的提升。

**致谢** 感谢陈栋华、崔辰州、梁助兴、林景明和张学军对本文的仔细审读及宝贵意见，同时感谢"星明天文台巡天群"成员对本文所作调查的积极参与。本文作者由GROWTH项目(National Science Foundation Grant No. 1545949)提供支持。

# Citizen Astronomy in China: Present and Future

Quan-Zhi Ye[1, 2]

[1] Astronomy Department, California Institute of Technology, Pasadena, CA 91125, U.S.A.
[2] Infrared Processing and Analysis Center, California Institute of Technology, Pasadena, CA 91125, U.S.A.

**Abstract:** Citizen science refers to scientific research conducted or participated by non-professional scientists (such as hobbyists or members from the general public). Citizen astronomy is a classic example of citizen science. Citizen astronomers has benefited from technological advancements in the recent decades as they fill the scientific gaps left by professional astronomers, in the areas such as time domain observations, visual classification and data mining. Chinese citizen astronomers have made a visible contribution in the discoveries of new objects; however, comparing to their counterparts in the western world, they appear to be less interested in researches that do not involve making new discovery, such as visual classification, long-term monitoring of objects, and data mining. From a questionnaire survey that aimed to investigate the motivation of Chinese citizen astronomers, we find that this population is predominantly male (92%) who mostly reside in economically developed provinces. A large fraction (69%) of the respondents are students and young professionals younger than the age of 25, which differs significantly from the occupation and age distribution of typical Chinese Internet users as well as the user distribution of large international citizen science projects such as the Galaxy Zoo. This suggests that youth generation in China is more willing to participate citizen astronomy research than average generation. Additionally, we find that interests in astronomy, desire to learn new knowledges, have a fun experience and meet new friends in the community are all important driving factors for Chinese citizen astronomers to participate research.

This also differs from their counterparts in western countries. With a large youth population that is interested in astronomy as well as a number of large astronomical facilities that are being planned or built, we believe that citizen astronomy in China has a vast potential. Timely and proper guidance from the professionals will be essential to help citizen astronomers to fulfill this potential.

**Key words:** sociology of astronomy

## 参考文献：

[1] Stebbins RA: Avocational science: The amateur routine in archaeology and astronomy [J]: International Journal of Comparative Sociology, 1980, 21: 34.

[2] Cunningham CJ and Orchiston W: The clash between William Herschel and the great German 'amateur' astronomer Johann Schroeter [B]: New Insights From Recent Studies in Historical Astronomy: Following in the Footsteps of F. Richard Stephenson, Springer, 2005, 205-222.

[3] Orchiston W and others: Amateur-professional collaboration in Australian science: the earliest astronomical groups and societies [J]: Historical records of Australian science, 1998, 12 (2): 163.

[4] Williams TR: Getting organized: A history of amateur astronomy in the United States [C]: Amateur - Professional Partnerships in Astronomy, ASP Conference Proceedings, Vol. 220. Edited by John R. Percy and Joseph B. Wilson, 2000, p. 3.

[5] Gada A, Stern AH, Williams TR: What Motivates Amateur Astronomers? [C]: Amateur - Professional Partnerships in Astronomy, ASP Conference Proceedings, Vol. 220. Edited by John R. Percy and Joseph B. Wilson, 2000, p. 14.

[6] Marshall PJ, Lintott CJ, Fletcher LN: Ideas for Citizen Science in Astronomy [J]: Annual Review of Astronomy and Astrophysics, 2015, 53: 247-278.

[7] Colgate SA, Moore EP, Carlson R: A fully automated digitally controlled 30-inch telescope [J]: Publications of the Astronomical Society of the Pacific, 1975, 87: 565-575.

[8] Castro-Tirado AJ: Robotic Autonomous Observatories: A Historical Perspective [J]: Advances in Astronomy, 2010, 570489.

[9] Cui CZ and others: Robotic Autonomous Observatory Network Review [J]: Progress in Astronomy (崔辰州等. 程控自主天文台网络的发展), 2013, 31(2): 141-159.

[10] Henze M and others: Supersoft X-rays reveal a classical nova in the M 31 globular cluster Bol 126 [J]: Astronomy & Astrophysics, 2013, 549, A120.

[11] Ye QZ and others: When comets get old: A synthesis of comet and meteor observations of the low activity comet 209P/LINEAR [J]: Icarus, 2016, 264, 48-61.

[12] Tan H and Gao X: 2017 BM3 [J]: Minor Planet Electronic Circulars, 2017, 2017-B51.

[13] Pastorello A and others: Massive stars exploding in a He-rich circumstellar medium - IX. SN 2014av, and characterization of Type Ibn SNe [J]: Monthly Notices of the Royal Astronomical Society, 2016, 456, 853-869.

[14] Lovejoy T and others: Comet C/2011 W3 (Lovejoy) [J]: International Astronomical Union Circulars, 2011, 9245.

[15] Kleyna JT and others: The Progressive Fragmentation of 332P/Ikeya-Murakami [J]: The Astrophysical Journal Letters, 2016, 827, L26.

[16] Tichy M and others: 2008 TC3 [J]: Minor Planet Electronic Circulars, 2008, 2008-T54.

[17] Ye QZ and others: Bangs and Meteors from the Quiet Comet 15P/Finlay [J]: The Astrophysical Journal, 2015, 814, 79.

[18] Samarasinha NH and others: Results from the worldwide coma morphology campaign for comet ISON (C/2012 S1) [J]: Planetary and Space Science, 2015, 118, 127-137.

[19] Liu JD: Rotation Period of 2019 van Albada [J]: Solar System Research, 2015, 44, 68-80.

[20] Liu JD: Rotation Period Analysis for 1967 Menzel [J]: Minor Planet Bulletin, 2016, 43, 98-99.

[21] Liu JD: Rotation Period Analysis for 2729 Urumqi [J]: Minor Planet Bulletin, 2016, 43, 204.

[22] Feng ZL and Xu PX: Reduction and Analysis of Visual Observations of η Aquarid Meteor Shower [J]: Acta Astronomica Sinica (冯占良，徐品新. 宝瓶η流星雨目视观测结果的归算和分析), 1990, 31 (3): 245-251.

[23] SonotaCo: Meteor shower catalog based on video observations in 2007-2008 [J]: WGN, Journal of the International Meteor Organization, 2009, 37 (2): 55-62.

[24] Cheng SH and Cheng SM: Meteor spectral observation with DSLR, normal lens and prism [J]: Meteor science, 2011, 39.

[25] Ogawa H and others: The 2002 Leonids as monitored by the International Project for Radio Meteor Observations [J]: WGN, Journal of the International Meteor Organization, 2002, 30: 225-231.

[26] Zhang YX and Zhao YH: Astronomy in the big data era [J]: Data Science Journal, 2015, 14.

[27] Westphal AJ and others: Stardust@ home: a massively distributed public search for interstellar dust in the stardust interstellar dust collector [C]: Lunar and Planetary Science XXXVI, Part 21, 2005.

[28] Raddick MJ and others: Galaxy Zoo: Motivations of citizen scientists [J]: Astronomy Education Review, 2013, 12 (1): 1-27.

[29] McMillan RS and others: The Spacewatch volunteer search for fast moving objects [J]: Minor Planet Bulletin, 2005, 32: 53.

[30] Liang ZX and others: Testing the rotating lighthouse model with the double pulsar system PSR J0737-3039A/B [J]: Monthly Notices of the Royal Astronomical Society, 2014, 439 (4): 3712-3718.

[31] Raddick MJ and others: Galaxy zoo: Exploring the motivations of citizen science volunteers [J]: Astronomy Education Review, 2013, 9 (1): 10103-10118.

[32] Price CA MJ and Paxson KB: The AAVSO 2011 Demographic and Background Survey [J]: Journal of the American Association of Variable Star Observers, 2012, 40: 1.